\documentstyle[preprint,eqsecnum,aps]{revtex}
\tighten
\begin{document}
\draft

\title {An exact self-consistent gravitational shock wave in
semiclassical gravity}

\author{Carlos O. Lousto\thanks{Electronic Address:
lousto@mail.physics.utah.edu}}
\address{Department of Physics, University of Utah\\
201 JBF, Salt Lake City, UT 84112, USA}

\author{Francisco D. Mazzitelli\thanks{Electronic Address:
fmazzi@df.uba.ar}}
\address{Departamento de F\'{\i}sica and IAFE\\ 
Facultad de Ciencias Exactas
y Naturales, Universidad de Buenos Aires\\ 
Ciudad Universitaria,
Pabell\'on I, (1428) Buenos Aires, Argentina}

\date{\today}
\maketitle
    
\begin{abstract}
We find a self-consistent pp-gravitational shock wave solution to the 
semiclassical Einstein equations resulting from the $1/N$ approach to
the effective action. We model
the renormalized matter stress-energy-momentum tensor by $N$ massless
scalar fields in the Minkowski vacuum plus a classical particle. We show
that quantum effects generate a milder singularity at the position of
the particle than the classical solution, but the singularity does not 
disappear. At large
distances from the particle, the quantum correction decreases slowly,
as $1/\rho^2$ ($\rho$ being the distance to the particle in the
shock wave plane). We argue that this large distance correction is
a necessary consequence of quantum gravity. 

\end{abstract}


\section{INTRODUCTION}

The backreaction of quantum fields on the spacetime geometry
can be  described by the so called Semiclassical Einstein Equation (SEE)
\cite{birrel}
\begin{equation}
G_{\mu\nu} + \alpha H_{\mu\nu} +\beta I_{\mu\nu}
+\Lambda g_{\mu\nu} =8\pi G\left(T_{\mu\nu}^{\text{class}}+
\langle T_{\mu\nu}\rangle \right) 
\label{bre}
\end{equation}
where one includes as a source the mean value of the energy momentum tensor
of the quantum fields. The higher derivative terms 
\begin{eqnarray}
H_{\mu\nu}&=&-2R_{;\mu\nu}+2g_{\mu\nu}\Box R-{1\over 2}g_{\mu\nu}
R^2+2RR_{\mu\nu}\nonumber\\
I_{\mu\nu}&=&-2R^{\alpha}_{\mu ;\nu\alpha}+\Box R_{\mu\nu}+
{1\over 2}g_{\mu\nu}\Box R+ 2R_{\mu}^{\alpha}R_{\alpha\nu}-
{1\over 2}g_{\mu\nu}R_{\alpha\beta}R^{\alpha\beta}
\label{hmunu}
\end{eqnarray}
are needed to renormalize the theory. The dressed coupling constants
$G,\Lambda,\alpha,\beta$ depend on an energy scale $\mu$ introduced 
by the regularization method. The dependence is given by
the renormalization group equations.
The mean value $\langle T_{\mu\nu}\rangle $ also depend on $\mu$ in such a way
that the full equation is $\mu$-independent.
The numerical values of $\alpha$ and $\beta$ should be determined
by experiments or could be computed from a more fundamental
theory. They should be of the order of $M_{Planck}^2$ for
$\mu\sim M_{Planck}$.  

Although a complete theory of quantum gravity from which the SEE
could be derived is still lacking, there are several arguments that
lead formally to this equation. On one hand, one can expand
the metric around a classical solution and quantize both the linearized 
metric fluctuations
and the matter fields on the classical background \cite{birrel}.
The one loop effective action for this field theory gives the leading
correction (in powers of $\hbar$) to the classical Einstein equation.
In this approximation, $\langle T_{\mu\nu}\rangle $ contains both graviton and
matter contributions. As two loop corrections will induce
additional terms proportional to $\hbar ^2$, in this context it
is reasonable to look only for solutions to the SEE which are 
perturbative in $\hbar$, and therefore close to the classical
solution. 
On the other hand, the SEE can be derived using a $1/N$ approximation.
Assume there are $N$ decoupled quantum matter fields. Taking the
limit $N\rightarrow\infty$ with $GN$ fixed (and rescaling also the
classical energy momentum tensor so that $GT_{\mu\nu}^{\text{class}}$ is
$N$-independent for large $N$), one arrives \cite{hartlehoro}
at an equation of the form (\ref{bre}). In this case, there is no
graviton contribution to $\langle T_{\mu\nu}\rangle $, since it is
suppressed by a factor $1/N$ with respect to the contribution
of the $N$ matter 
fields. Using this point of view, the SEE is valid to all orders
in $\hbar$, and it makes sense to look for exact, self-consistent
solutions. However, it should be kept in mind that some of the 
self-consistent solutions to the SEE may
be non-physical. 
It is not clear that one can ignore graviton effects at
Planck curvature, even for large $N$. Moreover, 
Eq.\ (\ref{bre}) contains time derivatives
of order higher than two, and therefore 
`runaway' solutions could arise. We refer the reader to
Ref.\cite{fw} for discussions on this important issue.

Physical, exact solutions to the SEE are interesting because they can show
important deviations from the classical behavior. The standard
lore about smearing of classical singularities due to quantum effects
should be proved in this context. The problem of the endpoint of
Hawking radiation should, and in fact it is, being analyzed using the SEE,
although   
progress has been important only in two dimensions \cite{strom}.
In four dimensions,  $\langle T_{\mu\nu}\rangle $ is a complicated, non-local
function of the geometry, and the SEE has been solved in a few 
cosmological cases\cite{Ch89}.
In this paper we are going to find a plane fronted with parallel rays
(pp) -- gravitational shock wave
self-consistent solution  to the SEE. From the exact solution we will 
be able to discuss the problem
of the smearing of the classical singularity. We will also compare
the exact and perturbative solutions. 

The paper is organized as follows.
In Section II we introduce the pp--gravitational shock waves,
and prove some useful geometric identities.
In Section III we use a resummation of the Schwinger DeWitt expansion
to write down the SEE for this particular metric. The SEE is solved
in Section IV. We discuss our results in Section V and leave for the
Appendix the case of the ultrarelativistic straight cosmic string.

\section{THE GRAVITATIONAL SHOCK WAVE}

It was stressed by 't Hooft\cite{tH87} that at Planckian energies
the relevant interaction among particles is the gravitational one,
which takes a particular simple form. In fact the gravitational
field generated by a highly energetic particle can be described
by the metric of a gravitational shock wave 

\begin{equation}
ds^2=-du~dv+f(x_\perp)\delta(u)~du^2+dx_\perp^2~,
\label{wave}
\end{equation}
where $u=t-z$ and $v=t+z$ are the usual null coordinates, and
$dx_\perp^2$ is the two--dimensional flat metric.
The wave propagates on a flat background
along the trajectory of the particle, i.e. $u=0$. The ``profile''
function $f(x_\perp)$ completely characterizes the spacetime.

Geodesics that cross the shock wave suffer a sudden shift in its $v$
coordinate given by\cite{DtH85}
\begin{equation}
\Delta v=-f(\rho_i)~,\label{salto}
\end{equation}
where $\rho_i$ is the distance to the inplunging geodesic from
the source on the surface $u=0$. This effect is represented
in Fig.\ \ref{figura1}.

Geodesics also have a refraction effect given by\cite{DtH85}
\begin{equation}
\cot(\theta_{in})+\cot(\theta_{re})=
{1\over2}\partial_\rho f(\rho_i)~,\label{refraccion}
\end{equation}
where $\theta_{in}$ and $\theta_{re}$ are the incident and refracted
angles respectively. This effect is depicted in Fig.\ \ref{figura2}.

The SEE (\ref{bre}) in principle involve terms 
containing powers of the curvature tensor, which in turn would be
proportional to powers of the delta-function $\delta (u)$. These
powers are meaningless in the context of classical distribution
theory. In order to avoid this problem, we will consider
a more general metric\cite{CL96}
\begin{equation}
ds^2=-du~dv+H(x_\perp,u)~du^2+dx_\perp^2~.
\label{genwave}
\end{equation}
We will show that for an arbitrary smooth function $H(x_\perp,u)$ the
SEE becomes linear in the curvature and, at the end of the calculation,
we will consider, the limit $H_n(x_\perp,u)\rightarrow f(x_\perp)\delta(u)$
($\{H_n\}$ is a convergent sequence of well-behaved test functions).

The curvature scalars of the metric\ (\ref{genwave})
formed out of the Ricci squared, Kretschmann,
and Curvature scalar tensors, all identically vanish
\begin{equation}
R\equiv0~,~~R_{\alpha\beta}R^{\alpha\beta}\equiv0~,~~
R_{\alpha\beta\gamma\delta}R^{\alpha\beta\gamma\delta}\equiv0~.
\label{escalares}\end{equation}
This does not mean, however, that the spacetime is free from any
singularity. In fact, the non-vanishing components of the
Riemann tensor are 
(apart from others obtained by symmetry properties)
\begin{equation}
R_{iuju}=-{1\over 2}\partial^2_{ij}H(x_{\perp},u)~,
\label{riemann}
\end{equation}
which may still diverge (as we will next see).

{}From Eq.\ (\ref{riemann}) we see that the only non-vanishing
component of the Ricci tensor is
\begin{equation}
R_{uu}=-{1\over 2}\nabla^2_{\perp}H~.
\label{ricci}
\end{equation}

Note that the only non vanishing components of the inverse metric
are $g^{uv}, g^{vu}, g^{ii}$ and $g^{vv}$. Also that the Christoffel
symbols with a subindex $v$,$\,\Gamma^{\mu}_{v \nu}$, vanish.
This implies that the covariant derivative $\nabla_v$ coincides
with the ordinary derivative when applied to a covariant index of
a tensor, and vanishes when acting on a $v$-independent function.

\bigskip

{}From the above properties it is easy to show that {\it any scalar
formed out of the Riemann  tensor and its derivatives must
vanish}. Indeed, a generic scalar can be built up as
\begin{equation}
g^{\mu\tau_1}g^{\nu\tau_2}...g^{..}
(\nabla_{\lambda_1}\nabla_{\lambda_2}...\nabla_
{\lambda_n}R_{\mu\nu\rho\sigma})(\nabla ...\nabla R...)...
(\nabla ...\nabla R...)~.
\label{scalar}
\end{equation}
The indices $\tau_i$ can be contracted with any of the covariant 
indices of the tensors in 
$(\nabla ...\nabla R...)$.
As the only non-vanishing components of the Riemann tensor
are those given by Eq.\ (\ref{riemann}), either $\mu$ or $\nu$
should equal $u$. Therefore, $\tau_1$ or $\tau_2$
should equal $v$. As $R_{v...}=0$ and $\nabla_v  R_{...} =0$,  
the scalar\ (\ref{scalar}) must vanish. Note that what we have
shown is that any expression containing $R_{\mu\nu\rho\sigma}$
vanishes if both indices $\mu\nu$ (or, by symmetry, $\rho\sigma$)
are contracted with any other index.

With similar arguments one can show that the most general
tensor with two covariant indices must be a linear combination of
$g_{\mu\nu}$ and $ F(\nabla^2_{\perp})R_{\mu\nu}$ [with $F$ an
arbitrary function].

The proof is as follows. Consider a generic tensor $(2,0)$
\begin{equation}
g^{\mu\tau_1}g^{\rho\tau_2}...g^{..}
(\nabla_{\lambda_1}\nabla_{\lambda_2}...\nabla_
{\lambda_n}
R_{\mu\nu\rho\sigma})
(\nabla ...\nabla R...)...(\nabla ...\nabla R...)
\label{tensor}
\end{equation}
The difference with respect to the scalar case is that now we can
have up to two non-contracted indices in the Riemann tensor.
If $R_{\mu\nu\rho\sigma}$ has three or four indices contracted,
the previous arguments imply that the tensor must vanish.
Therefore, $R_{\mu\nu\rho\sigma}$
must have exactly two non-contracted indices and,
as we are considering a $(2,0)$ tensor, it must be linear
in $R_{\mu\nu\rho\sigma}$.  For $\tau_1$ and $\tau_2$
there are two alternatives:

a) if $\tau_1$ is contracted with $\tau_2$, the tensor 
becomes $\nabla_{\lambda_1}\nabla_{\lambda_2}...
\nabla_{\lambda_n}
R_{\nu\sigma}$. 
The indices $\lambda_i$
must be contracted among themselves, and one finds something
proportional to $(\nabla^2_{\perp})^{{n\over 2}} R_{\nu\sigma}$.

b)   if $\tau_1$ and $\tau_2$ are contracted with
two derivatives $\lambda_1$ and $\lambda_2$, the result
is proportional to $\nabla^2_{\perp}R_{\nu\sigma}$. The
other derivatives must be contracted among themselves,
and one finds again  the same result.

\bigskip

The conclusion of this section is that any geometrical
tensor built up from the metric, with
two indices, and in particular, as we will soon see
$\langle T_{\mu\nu}\rangle $
(when the state is the  Minkowski's in-vacuum state), must be
a linear combination of $g_{\mu\nu}$ and a function
of $\nabla^2_{\perp}$ acting on $R_{\mu\nu}$. 

\section{THE SEMICLASSICAL EINSTEIN EQUATION}

Let us write Eq.\ (\ref{bre}) for the particular case of the
shock wave metric\ (\ref{wave}). Using the fact that $H_{\mu\nu}=0$,
$I_{\mu\nu}=\nabla^2_{\perp} R_{\mu\nu}$, we find
\begin{equation}
R_{\mu\nu}+\beta \nabla^2_{\perp} R_{\mu\nu}=8\pi G \left(
T_{\mu\nu}^{\text{class}}+ F(\nabla^2_{\perp}) R_{\mu\nu}\right)
\label{bresw}
\end{equation}
Here (as in Eq.\ (\ref{bre})),
we have split the stress--energy--momentum tensor into
two pieces: a classical source plus the renormalized expectation
value of the $T_{\mu\nu}$ due to $N$ quantum matter fields.
Since the state in which we take the expectation value is
Minkowski's vacuum, these fields do not contribute to the
classical source and, since the fluctuations of the classical source
are suppressed by a factor $1/N$ (as are the graviton contributions)
neither of these two terms appear in\ (\ref{bresw}) [For more details,
see, for example, Refs.\ \cite{C96,CL96b}.]

In general, we expect the function $F$ to depend on the mass
of the quantum fields and on the scale $\mu$ introduced 
by  the regularization method. For massive fields, this
function can be expanded in powers of $(\nabla^2_{\perp}/ m^2)$
(Schwinger--DeWitt expansion). For massless fields this expansion is 
inadequate, and $F$ becomes a non-analytic function of $\nabla^2_{\perp}$.
All this can be checked from the explicit calculations done
by Vilkovisky and collaborators\cite{BV87}.
They computed the effective action
up to quadratic terms in the curvature, and the energy momentum
tensor up to linear terms in the curvature. But for the 
pp-shock wave we have shown this is an exact result!
The energy momentum tensor
{\it is} linear in the curvature. Therefore,
the function $F$ is given by \cite{BV87,DM94}
\begin{equation}
F(\nabla^2_{\perp})=
{m^2\over 16\pi^2}(\xi - {1\over 6})\ln[{m^2\over\mu^2}]+
{\nabla^2_{\perp}\over 384\pi^2}
\int_0^1 dt\ t^4\ln\left[{m^2-{1\over 4}(1-t^2)\nabla^2_{\perp}
\over\mu^2}\right]~.
\label{formfactor}
\end{equation}
As anticipated, if $m^2\neq 0$, it is possible to expand $F$
in powers of $(\nabla^2_{\perp}/ m^2)$
\begin{equation}
F(\nabla^2_{\perp})=
{m^2\over 16\pi^2}(\xi - {1\over 6})\ln[{m^2\over\mu^2}]+
{\nabla^2_{\perp}\over 384\pi^2}
\int_0^1 dt\ t^4\left[\ln\left[{m^2\over\mu^2}\right]
-{1\over 4}(1-t^2){\nabla^2_{\perp}
\over m^2}+{\cal O}(({\nabla^2_{\perp}
\over m^2})^2)\right]~,
\label{formfactorapp}
\end{equation}
the Schwinger DeWitt expansion is recovered.

On the other hand, if $m^2=0$, the form factor
reads
\begin{equation}
F(\nabla^2_{\perp})={\nabla^2_{\perp}\over 1920\pi^2}
\ln\left[-{\nabla^2_{\perp}\over\mu^2}\right] + const\,\,\nabla^2_{\perp}
\label{formfactorm0}
\end{equation}
and it is a non-analytic function of the Laplacian.

It is interesting to note that in the massless case the form of the function
$F$ can be derived using general arguments
\cite{LM96}. Indeed, from dimensional
analysis we know that $F$ must be
of the form $F(\mu^2,\nabla^2_{\perp})
=\nabla^2_{\perp}{\tilde F}(\nabla^2_{\perp}/ \mu^2)$.
Moreover, $G$ is independent of $\mu$ and
${\beta}$ depends on $\mu$ according to the renormalization 
group equation \cite{DM94}
\begin{equation} 
{\mu\over 8\pi G}{d\beta\over d\mu}=-{1\over 960\pi^2}~.\label{running} 
\end{equation} 
As the whole semiclassical equation must be independent of $\mu$,
we must have $\mu d{\tilde F}/d\mu=-1/(960\pi^2)$. This
equation fixes the function $F$ up to a term that can be 
absorbed into a redefinition of $\beta$, and the result 
coincides with Eq.\ (\ref{formfactorm0}).

Now we are ready to write the differential equation for
the profile function. Assuming a point--like classical source
\footnote{This is an approximation, but if one wants to deal
with an extended source, our results below give the Green's function
for that case.} 
\begin{equation}
T_{uu}^{\text{class}}=p\delta(x_{\perp})\delta(u)
\end{equation}
we obtain
\begin{equation}
-{1\over 2}\left[\nabla^2_{\perp}+ 
\left(\beta(\mu) - {\cal A}\ln\left(-{\nabla_{\perp}^2\over
\mu^2}\right)\right)\nabla_{\perp}^4
\right]f(\rho)=8\pi Gp\delta(x_{\perp})~;~~{\cal A}={G\over 240\pi}
\label{profile}
\end{equation}
Note that the $\mu$-independence of the above equation is
explicit: Eq.(\ref{running}) implies that
the parenthesis multiplying $\nabla_{\perp}^4$ in Eq.\ (\ref{profile})
is independent of $\mu$. Eq.(\ref{profile}) has been solved
in Ref.\cite{CL96} for the case ${\cal A}=0$ (i.e. for the fourth order
theory).

\section{SOLUTION TO THE SEMICLASSICAL EQUATION}

Plane gravitational shock waves\ (\ref{wave}) have the nice property
of linearizing field equations. This feature allowed us to write
Eq.\ (\ref{profile}) in an explicit form. Again, this linearization
allow us to seek for the solution to the field equation by
using Fourier transform methods.
The result is
\begin{equation}
f(x_{\perp})= f_{GR} + 16\pi G p\int {d{\vec k}\over 4\pi^2}\,\,\,
e^{i{\vec k}\cdot{\vec x_{\perp}}}{\left[\beta -{\cal A}\ln
\left({k^2\over\mu^2}\right)\right ]\over
\left[1-\beta k^2+{\cal A}k^2\ln\left({k^2\over\mu^2}\right)\right]},
\label{sol}
\end{equation}
where $f_{GR}=-8Gp\ln(\rho/\rho_0)$ is the general relativistic
result \cite{AS71}, with $\rho_0$ an arbitrary constant. The 
second term in the above equation contains  the corrections
to the classical result.

To perform this integral we first note that since the problem has
axial symmetry it is convenient to pass from Cartesian to polar
coordinates in the perpendicular space, i.e. from $x_{\perp}$ to
$(\rho,\varphi)$. Doing so, the angular integral in Eq.\ (\ref{sol})
is a representation of the Bessel function\cite{GR80} $J_0$
\begin{equation}
f(\rho)= f_{GR}+ 8 G p\int_0^\infty dk\,\,\,
k\,J_0(k\rho){\left[\beta -{\cal A}\ln
\left({k^2\over\mu^2}\right)\right ]\over
\left[1-\beta k^2+{\cal A}k^2\ln
\left({k^2\over\mu^2}\right)\right]}.
\label{j0}
\end{equation}
Note that in the fourth order theory (${\cal A}=0,~\beta\neq 0$), 
the integrand
above has no real poles when the so called non-tachyon constraint,
$\beta <0$, is satisfied. When ${\cal A}\neq 0$, the condition
for absence of real poles is ${\cal A}\mu^2
\exp(\beta/{\cal A})<e$.

In this integral (now on the magnitude of ${\vec k}$), if ${\cal A}\neq 0$
we can absorb the term proportional to $\beta$ into a
redefinition of the scale $\mu$ 
\begin{equation}
f(\rho)= f_{GR}-8Gp\int_{0}^\infty
dk\,\,{{\cal A}\ln(k^2\mu^{-2}e^{-\beta/{\cal A}})
\over1+{\cal A}k^2\ln(k^2\mu^{-2}e^{-\beta/{\cal A}})}
\,k\,J_0(k\rho)~.
\label{bra}
\end{equation}

Let us now discuss some properties of the self-consistent
profile function. We first  note that it is finite at the
origin $\rho=0$. Indeed, we can write
\begin{eqnarray}
f(\rho)&=&-8pG\left[\ln(\rho /\rho_0)+K_0(\rho /\sqrt{\cal A})
-\int_0^\infty d\bar{k} J_0(\bar{k}\rho^*)\left({\bar{k}
\ln\left({\bar{k}^2\over{\cal A}^*}\right)\over
1+\bar{k}^2\ln\left({\bar{k}^2
\over{\cal A}^*}\right)}-{\bar{k}\over 1+\bar{k}^2}\right)\right]
\nonumber\\
&\dot=&f_{GR}+f_Q+\Delta f~.
\label{finite}
\end{eqnarray}
where we have taken dimensionless variables $\bar{k}\dot=\sqrt{\cal A}k$,
$\rho^*\dot=\rho/\sqrt{\cal A}$ and ${\cal A}^*\dot={\cal A}\mu^2
\exp(\beta/{\cal A})$ for ${\cal A}\not=0$. Also, we identified with
the subindices ``{\it GR}'', the general relativistic result, with
``{\it Q}'' the corrections induced by terms quadratic 
in the curvature (proportional to $K_0$), and 
with $\Delta f$ the higher order
corrections.

Using the asymptotic behavior of the modified Bessel function
$K_0$ and noting that
the last integral in Eq.\ (\ref {finite})
is finite in the limit $\rho\rightarrow 0$ one can easily
check that $f(\rho)$ is finite in this limit. This confirms 
the results of Ref. \cite{CL96} where the first two addends
of Eq.\ (\ref {finite}) have been obtained.

The next question is about the singularity at the origin.
Although all curvature scalars vanish for the pp-wave, the curvature 
tensor itself  may diverge. In fact,
\begin{equation}
R_{\rho u\rho}^v=\delta(u)\partial^2_\rho f(\rho)~,~~
R_{\phi uu}^\phi=-{\delta(u)\over2\rho}\partial_\rho f(\rho)~,~~
R_{\rho uu}^\rho={\delta(u)\over2}\partial^2_\rho f(\rho)~,
\end{equation}
that for the classical solution diverge as
$\partial^2_{\rho}f_{GR}\sim\rho^{-1}\partial_{\rho}f_{GR}\sim \rho^{-2}$.
When quadratic terms in the curvature are included in the classical
action, the divergence is milder, the components
of the curvature tensor diverge logarithmically \cite{CL96}
$\partial^2_{\rho}(f_{GR}+f_Q)\sim\rho^{-1}
\partial_{\rho}(f_{GR}+f_Q)\sim \ln\rho$.
We will now show that the divergence is still milder for the 
self-consistent
solution. The second derivative
of the profile function can be written as
\begin{equation}
\partial^2_{\rho}f=8 p G\int_0^{\infty} d\bar{k}
{\bar{k} J_0''(\bar{k}\rho^*)\over
1+\bar{k}^2 \ln{\bar{k}^2\over{\cal A}^*}}
\label{deriv}
\end{equation}
where the primes denote derivatives with respect to the argument.
Note that since $J_0''(0)=-1/2$, this integral diverges as the
$\ln[\ln()]$ of the upper limit.

To see the dependence with $\rho$ we split this integral as
\begin{equation}
\partial^2_{\rho}f
\simeq 8 p G\left[\int_0^{\bar{k}_0} d\bar{k}
{\bar{k} J_0''(\bar{k}\rho^*)\over1+\bar{k}^2 \ln{\bar{k}^2
\over{\cal A}^*}}+ \int_{\bar{k}_0}^{\infty} d\bar{k}
{J_0''(\bar{k}\rho^*)\over\bar{k}\ln{\bar{k}^2\over{\cal A}^*}}\right]
\label{derivsplit}
\end{equation}
where  $\bar{k}_0$ is such that $\bar{k}_0^2\ln{\bar{k}_0^2
\over{\cal A}^*}\gg1$.
The first term in Eq.\ (\ref{derivsplit}) is finite in the limit
$\rho\rightarrow 0$. The second term gives, after introducing
the new variable $y=\bar{k}\rho^*$ and performing an integration by parts
\begin{eqnarray}
\int_{\bar{k}_0\rho}^{\infty} {dy\over y}
{J''_0(y)\over \ln{y^2\over{\cal A}^*\rho^{*2}}}&=&{1\over2}
J_0''(y)\ln\left[\ln\left({y^2\over{\cal A}^*\rho^{*2}}\right)\right]
\vert_{\bar{k}_0\rho^*}^{\infty}
-{1\over 2}\int_{\bar{k}_0\rho^*}^{\infty}dy\,\,J'''(y)
\ln\left[\ln\left({y^2\over{\cal A}^*\rho^{*2}}\right)\right]
\nonumber\\ \nonumber\\
&\simeq&{1\over 2}J''(\bar{k}_0\rho^*)\left\{\ln[\ln({1\over {\cal A}^*
\rho^{*2}})]-\ln[\ln(\bar{k}_0/{\cal A}^*)]\right\}~.
\label{f''app}
\end{eqnarray}
Using that $J''(0)=-1/2$,
we see that $\partial^2_{\rho}f\simeq 2pG\ln[-\ln(\rho^{*})]$
as $\rho\rightarrow 0$. A similar result
can be shown for $\rho^{-1}\partial_{\rho}f$. The conclusion
is that the curvature diverges like $\ln[-\ln\rho^{*}]$
near the position of the particle.

What happens at large distances? In this case we expect the self-consistent
solution to approach the solution perturbative in $\hbar$ (perturbative
in ${\cal A}$, in our notation). From Eq.\ (\ref{finite}) we see 
that, to lowest order in ${\cal A}$ (for large $\rho$),
\begin{equation}
\Delta f(\rho)\simeq -8pG\int_0^{\infty}d\bar{k}\,\,\, 
\bar{k}J_0(\bar{k}\rho^*)\ln\left({\bar{k}^2\over{\cal A}^*}\right)~.
\label{fapprox}
\end{equation}
This integral is ill defined and must be treated as a distribution.
It is essentially the Fourier transform of $\ln(\bar{k}^2/{\cal A}^*)$
in two dimensions. To compute it, we introduce an integral representation
for the logarithm
\begin{equation}
\Delta f(\rho)\simeq 8Gp\int_0^{\infty}dz\int_0^{\infty}d\bar{k}\,\,
\bar{k} J_0(\bar{k}\rho^*)\left[{1\over z+\bar{k}^2}-{1\over z+
{\cal A}^*}\right]~.\label{repr}
\end{equation}
The ${\cal A}^*-$dependent term gives a contribution proportional to
$\delta(x_{\perp})$ which is irrelevant at large distances. Omitting
this term we obtain
\begin{equation}
\Delta f(\rho)\simeq8Gp\int_0^{\infty}dz\,\,K_0(\sqrt z \rho^*)
={16Gp{\cal A}\over\rho^2}\,\, .
\end{equation}
Therefore, we expect that, at large distances,
$f(\rho)=f_{GR}+16Gp{\cal A}/\rho^2$.
It is important to stress that the 
perturbative and self-consistent solutions are completely different
unless $\rho\rightarrow\infty$.

We have numerically computed the self-consistent solution
starting from Eq.\ (\ref{finite}), and
confirmed the properties of the profile
function both at the $\rho\rightarrow 0$ and $\rho\rightarrow\infty$
limits. The results for $\Delta f$ and $\partial^2_{\rho}f$
are shown in Figs.\ (\ref{figura3}) and (\ref{figura4}) .
In those figures we only show  curves for ${\cal A}^*<e$, since
$\Delta f$ diverges for ${\cal A}^*=e$. This is due to
the fact that, as we already pointed out, the integrand
in Eq. (\ref{sol}) develops a double real pole as ${\cal A}^*$ approaches
$e$, that becomes two single real poles for ${\cal A}^*> e$. These
two real poles, $k_p$, change the asymptotic behavior of $\Delta f$
for large $\rho^*$, making it to decrease slowly (like 
a linear combination of $J_0(k_p\rho), \,\,p=1,2$).
However, from the definition of ${\cal A}^*$ right after Eq.\ (\ref{finite}),
we and the renormalization group equation\ (\ref{running}), we see that
${\cal A}^*$ is actually independent of $\mu$. Then, from the renormalized
value of ${\cal A}$, given in Eq.\ (\ref{profile}),
and the no tachyon constraint
that implies $\beta<0$, we can infer that ${\cal A}^*\ll1$.

\section{Discussion}

We have found a self-consistent pp-shock wave solution to
the SEE in the $1/N$ approximation.

We have shown that quantum effects make milder the singularity
of the classical solution at the origin. In the classical
theory, the curvature diverges like $1/\rho^2$. 
In the quadratic theory it diverges like $\ln(\rho^2)$. When the
backreaction effects of massless quantum scalar fields
are taken into account, it diverges like $\ln[-\ln(\rho^2)]$. 
Therefore, the 
self-consistent solution can be trusted up to a distance very
close to the position of the particle, where the curvature
becomes of the order of Planck curvature.

At large distances, the situation is similar to the case of
the corrections to the Newtonian potential: while classical
terms quadratic in the curvature produce an exponentially
decreasing correction\cite{stelle}
\begin{equation}
V=-{GM\over r}(1-\exp(-2\sigma r))~,~~\sigma^2=3\alpha+\beta~,
\label{newt}
\end{equation}
backreaction effects due to massless quantum
fields induce a power law correction of the form  \cite{DM94,D94}
\begin{equation}
V=-{GM\over r}+ {aG^2M\over r^3}\,.
\label{newtq}
\end{equation}
This is the dominant correction at large distances. The numerical
constant $a$ depends on the matter content of the theory and 
includes a contribution from gravitons.

We have a similar picture for the pp-gravitational shock wave. The
profile function in the quadratic theory is given by \cite{CL96}
\begin{equation}
f_{GR}+f_Q=f_{GR}(\rho)-
8GpK_0(\rho/\sqrt{-\beta})\,,
\end{equation}
and the correction to the GR result is exponentially small
at large distances. However, we have shown that quantum massless
fields induce a correction of the form
\begin{equation}
f(\rho)\simeq f_{GR}(\rho)+{16Gp{\cal A}\over \rho^2}~;~~\rho\to\infty~,
\end{equation}
which again is the leading correction at large distances.

If one treats General Relativity as an effective field theory of an
unknown quantum theory of gravity,  
it can be shown that
the leading quantum
corrections in Eq.\ (\ref{newtq}) will not depend on the details
of the more fundamental theory: they are a consequence of quantum gravity,
whatever this theory may be \cite{D94}. The same argument applies
to the profile function of the gravitational shock wave: we expect
quantum gravity effects to produce large distance corrections 
proportional to ${\cal C}/\rho^2$. The coefficient ${\cal C}$
should be determined by a more careful calculation that should
take into account all massless fields in nature
including graviton contributions to the renormalized
energy--momentum tensor. However, this contribution should be
qualitatively similar to the one computed for massless scalar fields
in this paper.

Since we are considering exact solutions to the semiclassical 
Einstein equations\ (\ref{bre}), it is worth discussing here 
the question of the runaway solutions that such higher order
equations may have. In Ref.\cite{PS93} it was proposed that a
way of getting rid of such unwanted runaway solutions is to only
consider first order perturbative solutions in the corrections
to general relativity
(here proportional to ${\cal A}$). This is clearly too restrictive
in the framework presented in our paper.
In the $1/N$ approach ($N\to\infty$) the semiclassical Einstein 
equations hold exactly. In fact, since the scalar fields are taken to 
be free (i.e. non-self-interacting) only the leading order terms in 
these scalar fields arise. Consequently, one is not only allowed 
to go further than the first order perturbative regime, but one should 
also take into account all the ``well-behaved'' exact solutions 
to the semiclassical fields equations. The well--behaved
solutions we found have been selected from all possible ones
by allowing them to be represented by a Fourier transform.

\begin{acknowledgments}
C.O.L. was partially supported by NSF Grant No. PHY-95-07719
and by research funds of the University of Utah. F.D.M. 
was supported by Universidad de Buenos Aires, CONICET and
Fundaci\' on Antorchas. The authors thank M. Campanelli for useful
discussions.
\end{acknowledgments}

\appendix
\section{Ultrarelativistic cosmic string}

It is also of interest to consider the case of an ultrarelativistic
particle in $2+1$ dimensions, since its results are also valid
for a straight cosmic string boosted in a direction perpendicular
to it\cite{LS91,CL96}. 

In this case the field equation simply takes the form of
Eq.\ (\ref{profile}) with $\nabla_{\perp}^2$ replaced by
$\partial_y^2$, where $y$
is the Cartesian coordinate perpendicular to both the direction of
motion and the axis along the string.

Consequently, the Fourier decomposition of the profile function
leads to
\begin{equation}
f(|y|)=f_{GR}(|y|)+ 
16\pi G \tilde{p}\int_{-\infty}^\infty {d{k}\over 2\pi}\,\,\,
e^{ik|y|}{\left[\beta -{\cal A}\ln
\left({k^2\over\mu^2}\right)\right ]\over
\left[1-\beta k^2+{\cal A}k^2\ln\left({k^2\over\mu^2}\right)\right]},
\label{soly}
\end{equation}
where now $\tilde{p}$ represents the momentum per unit length of
the cosmic string and $f_{GR}=-8\pi\tilde {p}|y|$ is the general relativistic
result found in Ref.\ \cite{LS91}.

By splitting  the integral in Eq.(\ref{soly}) we obtain
\begin{eqnarray}
f(|y|)&=&f_{GR}(|y|)+ 8G \tilde{p}\Biggr\{-2\sqrt{\cal A}\int_0^
\infty d\bar{k}
{\cos(\bar{k}|y|^*)\over 1+\bar{k}^2}\nonumber\\
&&+2\sqrt{\cal A}\int_0^\infty d\bar{k}\cos(\bar{k}|y|^*)
\left[1-\ln(\bar{k}^2/{\cal A}^*)\over(1+\bar{k}^2)
(1+\bar{k}^2\ln(\bar{k}^2/{\cal A}^*))\right]\Biggr\}~,
\end{eqnarray}
where the dimensionless variables have the same meaning as given
after Eq.\ (\ref{finite}). The first integral can be found
in tables\ \cite{GR80} and its result is
$\pi\exp\{-|y|/\sqrt{\cal A}\}/2$. Thus giving the quadratic
theory correction as found in Ref.\ \cite{CL96}. The second integral
has to be performed numerically as we did before for the particle
in four dimensions.

It is easy to check that
\begin{equation}
f''(|y|)={-16G \tilde{p}\over\sqrt{\cal A}}\int_0^\infty d\bar{k}\,\,\,
{\cos(\bar{k}|y|^*)\over 1+\bar{k}^2\ln(\bar{k}^2/{\cal A}^*)}~,
\end{equation}
is finite at $|y|=0$. The Riemann tensor is then also finite
(apart from the $\delta(u)$ behavior). Note that this is already true
in the quadratic theory; while in general relativity, the divergence
was as $\delta(|y|)$.

The behavior for large $|y|$ can be computed on the same lines as before,
and we obtain
\begin{equation}
f(|y|)\simeq -8\pi G\tilde{p}|y|+{16\pi G\tilde{p}{\cal A}\over|y|}~;~~
 |y|\to\infty~.
\end{equation}
Once more, we see that quantum effects due to massless fields are the
leading corrections at large distances.

\begin{figure}
\caption{The dashed line represents the path of a null geodesics in the 
$(u, v)$ plane. Since for $u\neq 0$ the spacetime is flat, geodesics are 
straight lines. At $u=0$ they have a discontinuity jump described by 
Eq.\ (\protect\ref{salto}). The point--like source $p$ is located at
the origin and the plane $u=0$ contains the gravitational shock wave.}
\label{figura1}
\end{figure}
\begin{figure}
\caption{The `spatial refraction' of a null geodesics as described
by Eq.\ (\protect\ref{refraccion}). Here $\theta_{in}$ and
$\theta_{re}$ are the incident and refracted angles respectively.}
\label{figura2}
\end{figure}
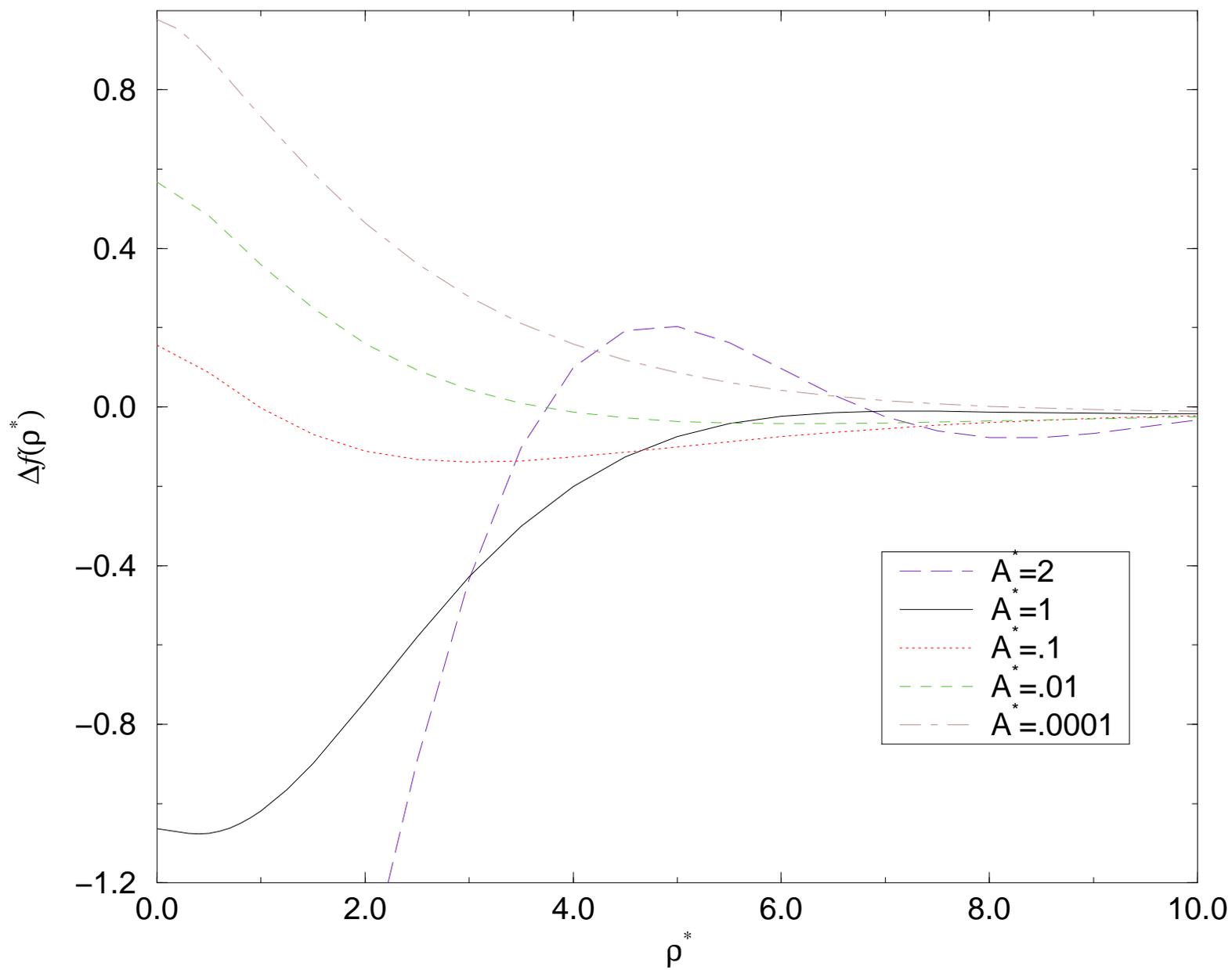
\begin{figure}
\caption{The higher order corrections to the profile function of the
shock wave as defined in Eq.\ (\protect\ref{finite}) in units of $8pG$.
Here $\rho^*=\rho/\protect\sqrt{\cal A}\protect\cong
27.5l_{Pl}^{-1}\rho$ and
${\cal A}^*\dot={\cal A}\mu^2\exp(\beta/{\cal A})$ are convenient
dimensionless variables. The correction reaches a finite value at
$\rho=0$ and vanishes as $1/\rho^{*2}$ for large $\rho^*$.}
\label{figura3}
\end{figure}
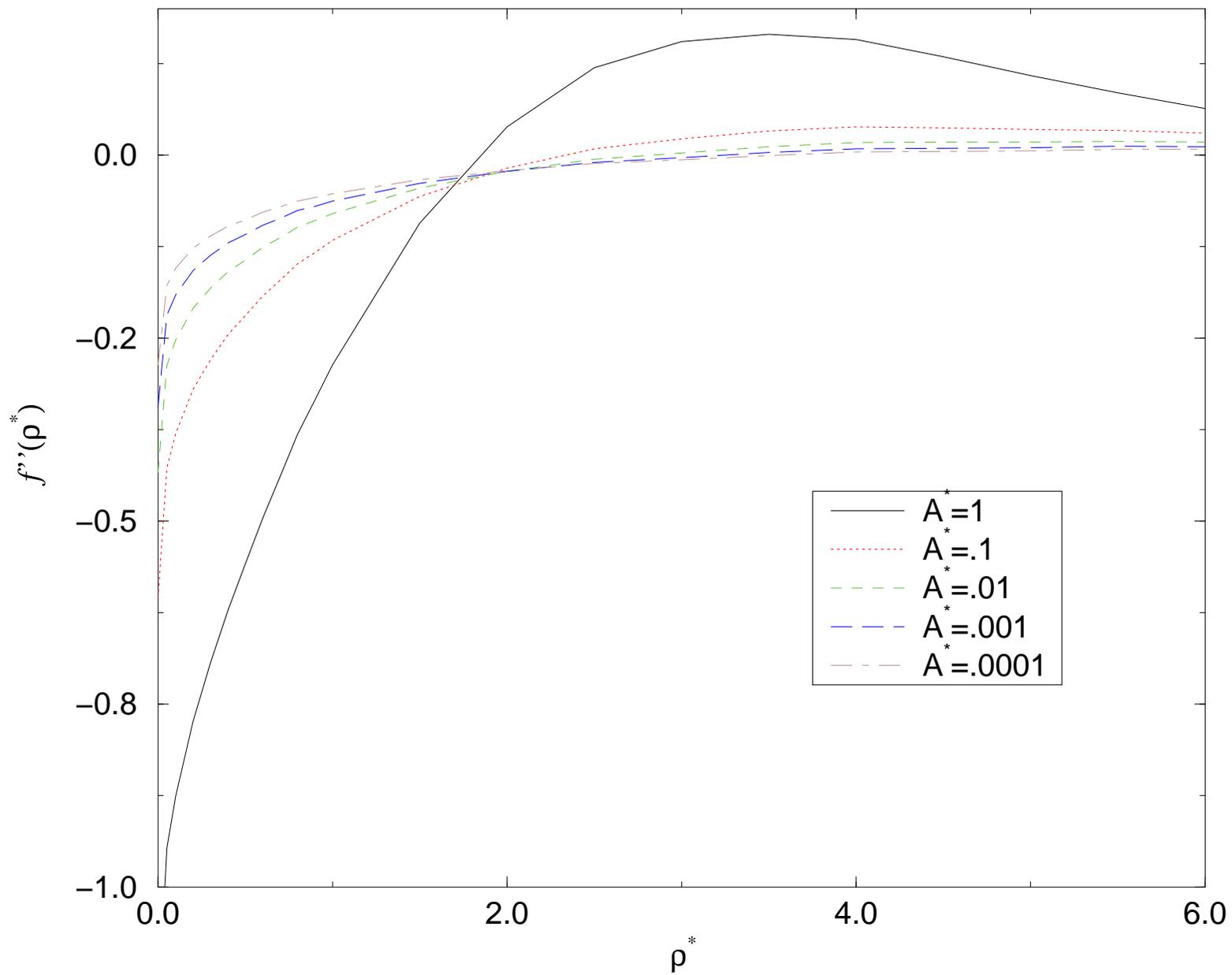
\begin{figure}
\caption{$d^2f/d\rho^2$ as defined in Eq.\ (\protect\ref{deriv})
over $8pG$.
This is proportional to the non-vanishing Riemann tensor components
and show that they diverge as $\ln[\ln(\rho^*)]$ for $\rho^*\to0$.}
\label{figura4}
\end{figure}


\begin{references}

\bibitem{birrel} N. D. Birrell and P. C. W. Davies, {\it Quantum Fields
in Curved Space}, Cambridge University Press, Cambridge (1982).\par

\bibitem{hartlehoro} J. B. Hartle and G. T. Horowitz, Phys. Rev.
D {\bf 24}, 257 (1981).\par

\bibitem{fw} E. E. Flanagan and R. M. Wald, Phys. Rev. D {\bf 54},
6233 (1996), and references therein.\par

\bibitem{strom} A. Strominger, "Les Houches Lectures on Black
Holes", hep-th 9501171, and references therein.\par

\bibitem{Ch89} M. V. Fischetti, J. B. Hartle
and B. L. Hu, Phys. Rev. D {\bf 20}, 1757 (1979); J.B. Hartle and
B. L. Hu, ibid 1772 (1979); A. A. Starobinsky, Phys. Lett. {\bf B}
91, 99 (1980); P. Anderson, Phys. Rev. D {\bf 29}, 615 (1984).\par


\bibitem{tH87}G. 't Hooft, Phys. Lett. B {\bf 198}, 61 (1987).\par

\bibitem{DtH85}T. Dray and G. 't Hooft, Nucl. Phys. B {\bf 253}, 173
(1985).

\bibitem{C96} M. Campanelli, Ph. D. Thesis, University of Bern,
June 1996 (unpublished).

\bibitem{CL96b} M. Campanelli and C. O. Lousto, preprint gr-qc/9608027.

\bibitem{BV87} A. O. Barvinsky and G. A. Vilkovisky, Nucl. Phys. B
{\bf 282}, 163 (1987); ibid B {\bf 333}, 471 (1990); G. A. 
Vilkovisky in {\it Quantum Theory of Gravity}, ed. by S.M.
Christensen (Hilger Bristol, 1984.\par

\bibitem{LM96} F.C. Lombardo and F.D. Mazzitelli, preprint gr-qc/9609073.

\bibitem{DM94}D. A. R. Dalvit and F. D. Mazzitelli,
Phys. Rev. D {\bf  50}, 1001 (1994).\par

\bibitem{CL96} M. Campanelli and C. O. Lousto,
Phys. Rev. D {\bf 54}, 3854 (1996).


\bibitem{AS71}P. C. Aichelburg and R. U. Sexl, Gen. Rel. Grav.
{\bf 2}, 303 (1971).

\bibitem{GR80} I. S. Gradshteyn and I. M. Ryzhik,
{\it Tables of integrals, Series and Products},
Academic Press, New York (1980).


\bibitem{stelle} K.S. Stelle, Gen. Rel. and Grav. {\bf 9}, 353 (1978).\par

\bibitem{D94} J.F. Donoghue, Phys. Rev. D {\bf 50}, 3874 (1994).\par

\bibitem{PS93} L. Parker and J. Z. Simon, Phys. Rev. D, {\bf 47},
1339 (1993).\par

\bibitem{LS91} C. O. Lousto and N. S\'anchez, Nucl. Phys. B {\bf 355},
231 (1991).\par



\end{references}
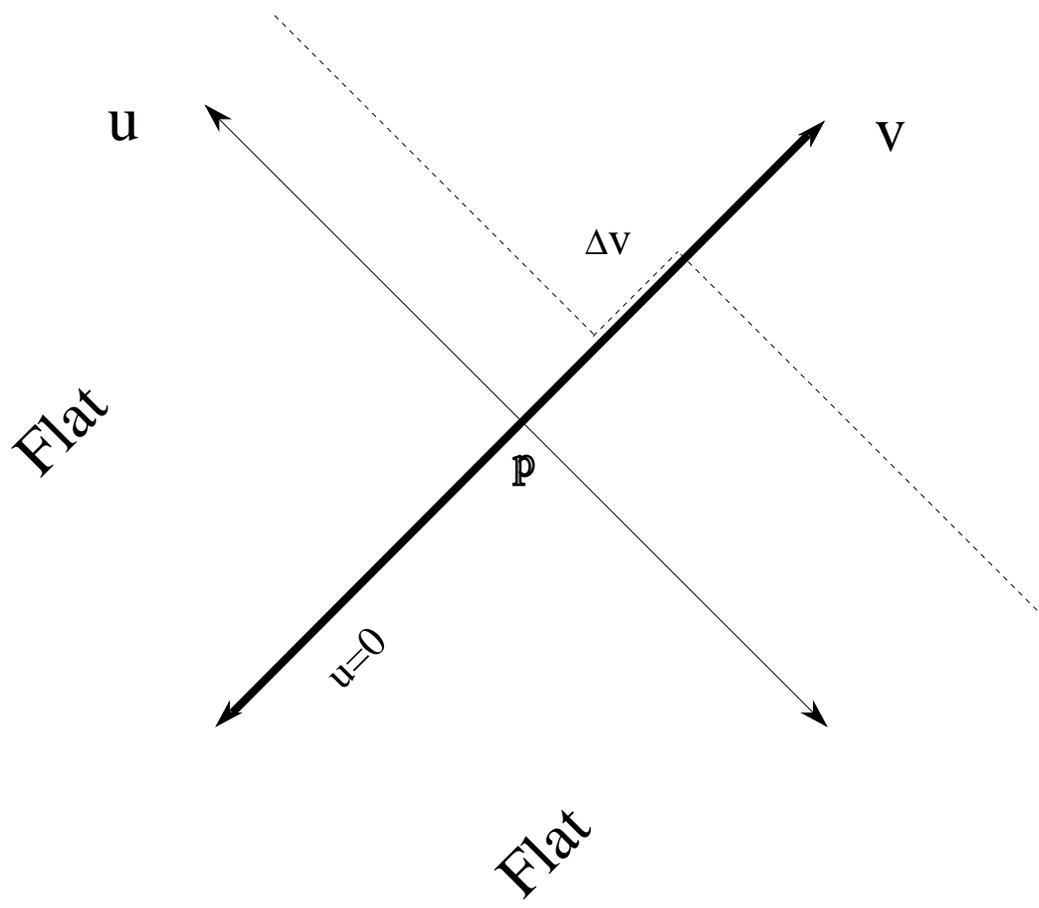
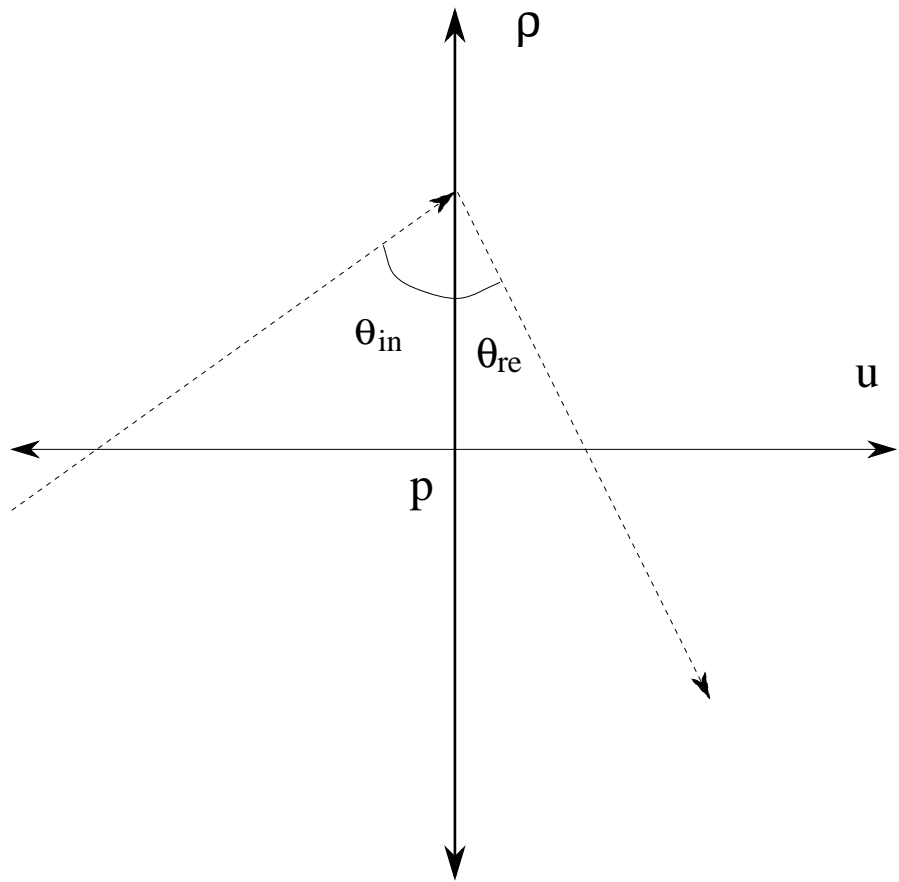
\end{document}